\documentclass{scrartcl}

\usepackage{amsmath}
\usepackage{amssymb}

\usepackage{graphicx}

\usepackage{hyperref}
\usepackage{doi}
\usepackage[numbers]{natbib}
\usepackage{float}

\usepackage{color} 

\usepackage{setspace} 
\let\d\relax
\DeclareMathOperator{\d}{d}

\begin{document}

% Title must be 150 characters or less
\begin{flushleft}
{\Large
\textbf{Pattern of tick aggregation on mice: larger than expected distribution tail enhances the spread of tick-borne pathogens}
}
% Insert Author names, affiliations and corresponding author email.
\\
Luca Ferreri$^{1,2,\ast}$, Mario Giacobini$^{1,2,3}$, Paolo Bajardi$^{1,2}$, Luigi Bertolotti$^{1}$, Luca Bolzoni$^{4,5}$, Valentina Tagliapietra$^{5}$, Annapaola Rizzoli$^{5}$, Roberto Ros\`a$^{5}$\\
\bf{1} Computational Epidemiology Group, Department of Veterinary Sciences, University of Torino, Italy
\\
\bf{2} Applied Research on Computational Complex Systems Group, Department of Computer Science, University of Torino, Italy
\\
\bf{3} Complex Systems Unit, Molecular Biotechnology Centre, University of Torino, Italy
\\
\bf{4} Istituto Zooprofilattico Sperimentale della Lombardia e dell'Emilia Romagna, Parma, Italy
\\
\bf{5} Dipartimento Biodiversit\`a ed Ecologia Molecolare, Centro Ricerca e Innovazione, Fondazione Edmund Mach, San Michele all'Adige (TN), Italy
\\
$\ast$ E-mail: \href{mailto:luca.ferreri@unito.it}{luca.ferreri@unito.it}
\end{flushleft}

\section*{Abstract}

The spread of tick-borne pathogens represents an important threat to human and animal health in many parts of Eurasia. Here, we analysed a 9-year time series of \textit{Ixodes ricinus} ticks feeding on \textit{Apodemus flavicollis} mice (main reservoir-competent host for tick-borne encephalitis, TBE) sampled in Trentino (Northern Italy). The tail of the distribution of the number of ticks per host was fitted by three theoretical distributions: Negative Binomial (NB), Poisson-LogNormal (PoiLN), and Power-Law (PL). The fit with theoretical distributions indicated that the tail of the tick infestation pattern on mice is better described by the PL distribution. Moreover, we found that the tail of the distribution significantly changes with seasonal variations in host abundance. In order to investigate the effect of different tails of tick distribution on the invasion of a non-systemically transmitted pathogen, we simulated the transmission of a TBE-like virus between susceptible and infective ticks using a stochastic model. Model simulations indicated different outcomes of disease spreading when considering different distribution laws of ticks among hosts. Specifically, we found that the epidemic threshold and the prevalence equilibria obtained in epidemiological simulations with PL distribution are a good approximation of those observed in simulations feed by the empirical distribution. Moreover, we also found that the epidemic threshold for disease invasion was lower when considering the seasonal variation of tick aggregation. 
% Please keep the Author Summary between 150 and 200 words
% Use first person. PLoS ONE authors please skip this step. 
% Author Summary not valid for PLoS ONE submissions.   

\section*{Author Summary}

Our work analyses a 9-year time series of tick co-feeding patterns on Yellow-necked mice. Our data shows a strong heterogeneity, where most mice are parasitised by a small number of ticks while few host a much larger number. We describe the number of ticks per host by the commonly used Negative Binomial model, by the Poisson-LogNormal model, and we propose the Power Law model as an alternative. In our data, the last model seems to better describe the strong heterogeneity. In order to understand the epidemiological consequences,  we use a computational model to reproduce a peculiar way of transmission, observed in some cases in nature, where uninfected ticks acquire an infection by feeding on a host where infected ticks are present, without any remarkable epidemiological involvement of the host itself. In particular, we are interested in determining the conditions leading to pathogen spread. We observe that the effective transmission of this infection in nature is highly dependent on the capability of the implemented model to describe the tick burden. In addition, we also consider seasonal changes in tick aggregation on mice, showing its influence on the spread of the infection.
% Results and Discussion can be combined.

\section*{Introduction}

Several ecological studies have shown that the distribution of ticks on their hosts is often highly aggregated, with a large number of hosts harbouring few parasites and a small number harbouring a large number of them (\cite{Crofton1958,Crofton1971,Plowright1977,AM1978,MA1978}; other interesting references could be found in \cite{Shaw1995}). In addition,  the distribution of tick development stages is coincident, rather than independent \cite{Perkins2003}. Specifically, those hosts feeding larval tick stages were simultaneously feeding the greatest number of nymphs. As a result, about $20\%$ of all hosts feed $80\%$ of both larvae and nymphs and the number of larvae feeding alongside nymphs is twice as many as it would be if the distributions were independent \cite{Randolph1999,Perkins2003}. The aggregation of parasites on hosts bears important implications for vector-borne disease dynamics, since the small fraction of hosts supporting the bulk of the vector population is also responsible for the majority of the pathogen transmission \cite{Woolhouse1997}.

The transmission of tick-borne diseases is characterised by an intricate set of ecological and epidemiological relationships between pathogen, tick vector, vertebrate hosts and humans that largely determine their temporal and spatial dynamics \cite{Randolph2010}. Tick-borne disease dynamics feature several complexities, due to the presence of a number of heterogeneities in the system coupled with non-linear phenomena operating in the transmission processes between ticks, host and pathogen \cite{Randolph2002}. The transmission of pathogens from one tick to another, a pre-requisite for the establishment of cycles of infection, may occur via three different pathways depending on the pathogen (see \cite{Randolph1998} for a comprehensive review). First, adult female ticks may transmit the pathogen to eggs trans-ovarially. Second, ticks may infect a host during their blood meal, leading to a systemic infection in the host; ticks might then acquire the infection by feeding on an infected host, maintaining the infection trans-stadially. Third, ticks may become infected by co-feeding with infected ticks on the same host. Co-feeding transmission is also called non-systemic as it does not require the host to have a systemic infection, since pathogens are transmitted from one tick to another as they feed in close proximity. Vertebrate hosts may vary in their competency to support systemic and co-feeding transmission \cite{Mannelli2012}. Tick-borne pathogens differ also for the mechanisms which they use to persist in nature. For instance, \textit{Rickettsia} spp., the pathogen agents causing Rocky Mountain Spotted Fever, are maintained by systemic and trans-ovarial transmission in \textit{Dermacentor variabili} and \textit{andersoni} \cite{Amsden2005} while it has been observed that \textit{Borrelia Burgdoferi} s.l. spirochaetes persist in nature by taking advantage of all three routes of transmission in \textit{I. ricinus}, \cite{lyme,Mannelli2012}.

In the case of the tick-borne encephalitis virus (TBEv), which is an increasing public health concern in Europe \cite{RandolphSumilo2007,Sumilo2007,Stanek09}, trans-ovarial transmission seems to be relatively rare and its contribution is generally thought to be negligible \cite{Nuttall2003}. On the other hand, both systemic and non-systemic transmission can take place on reservoir-competent rodent hosts. However, due to the very short duration of the TBEv infection in rodents, \cite{Randolph1996}, the systemic route would only allow infection of a very limited number of ticks. Indeed, non-systemic transmission through co-feeding ticks is a more efficient transmission route for TBE \cite{Randolph1996,Randolph1999}. Different studies have shown that TBEv would not become established in competent hosts, such as rodents, without the amplification of the overall transmission efficiency provided by co-feeding transmission (see for instance \cite{Randolph1996,Rosa2007b,Hartemink2008,Harrison2012}). The aggregation pattern of ticks on hosts therefore plays a more important role in the transmission of TBEv than in other tick-borne pathogens, such as \textit{Borrelia burgdorferi} sensu lato and \textit{Anaplasma phagocytophilum,} where other efficient routes of transmission have been observed.

Tick aggregation on hosts and correlation of tick stages facilitate co-feeding transmission and thus significantly increase the basic reproductive number, $R_{0}$, of the pathogen, with direct implications for its persistence \cite{Rosa2003,Harrison2012}. Using different levels of aggregation (from independent to coincident aggregated distribution), Harrison and collaborators \cite{Harrison2012} showed that values of $R_{0}$ increase with progressive levels of aggregation, making it more likely for tick-borne pathogens to become established and persist. In addition, the authors of the cited works evinced that when ticks followed a coincident aggregated distribution, the increase of $R_{0}$ was greater than in the case of independent aggregated distributions.

The degree of aggregation of ticks can be measured in a number of ways. Since the appearance of influential works by Randolph \cite{Randolph1975} and Shaw et al. (\cite{Shaw1995} and \cite{Shaw1998}) the negative binomial (NB) distribution has been extensively used to describe tick aggregation on hosts (see e.g. \cite{Brunner2008,Kiffner2011,Harrison2012}). Alternatively, other works suggested that different distributions characterised by larger tails than NB (i.e., predicting more rodents with very large tick burden than expected with NB), can be effective in describing tick aggregations. Specifically, a Poisson-LogNormal (PoiLN) mixed model has been successfully used to describe tick distribution on red grouse chicks \cite{Elston2001}, while Bisanzio and collaborators \cite{Bisanzio2010} showed the first evidence that the distribution heterogeneity of ticks on hosts seemed to be better described by a power-law (PL) than a negative binomial distribution. A suitable description of the distribution tail might have important consequences on the dynamics of the pathogen spreading process. Modelling the spread of vector-borne diseases through bipartite networks \cite{Bisanzio2010} showed that the extreme aggregation of ticks on hosts has dramatic consequences on the behaviour of the epidemic threshold.

In the current study we used an extensive data set of \textit{Ixodes ricinus} ticks feeding on mice (a total of 4722 parasitised hosts collected in 9 years) to detect the best fit for the distribution of tick burden on mice by testing the performance of NB and PoiLN versus PL distribution, with particular interest in the shape of the distribution tail which is crucial to suitably describe the fraction of co-feeding ticks necessary for TBEv transmission. Then, we used a stochastic model to simulate the effect of fitting different tick distributions on the infection dynamics of a tick-borne pathogen. Specifically, we investigated the spread of a non-systemically transmitted pathogen (e.g. TBEv) by modelling the pathogen transmission between susceptible and infective ticks, considering only co-feeding transmission and distributing ticks on mice under the hypotheses of NB, PoiLN, and PL distributions. Finally, we investigated the seasonal variations in the pattern of tick burden distribution on mice and its implication on TBE-like infection dynamics.

%% ==================
%% ==================

\section*{Materials and Methods}
\subsection*{Ethics Statement}
All animal handling procedures and ethical issues were approved by the Provincial Wildlife Management Committee (renewed authorisation n. 595 issued on 04.05.2011)

\subsection*{Tick Burden Data}

Rodent tick burden data was collected by trapping mice using capture-marking-recapture techniques during 2000-2008. The study area was a mixed broadleaf woodland \cite{Perkins2003,Rosa2007b}, located in Valle dei Laghi within the Autonomous Province of Trento, in the north-eastern Italian Alps (grid reference 1652050E 5093750N, altitude 750-800 m a.s.l.). In the year 2000, mice were monitored in nine selected areas through placement of 8x8 trapping grids with a 15-m inter-trap interval. In 2001 and 2002 the number of trapping grids was reduced to eight, while from 2003 onward their number was further reduced to four.

In summary, the trapping effort consisted of $129$ twice-daily trap sessions with at least one capture, resulting in a total number of $4722$ \textit{Apodemus flavicollis} captured with at least one tick attached. For each captured rodent the number and life stage of feeding ticks was carefully assessed and registered, without removal \cite{Perkins2003,Rosa2007b}. A total number of $55411$ ticks were counted of which $98.64\%$ were larvae, $1.30\%$ were nymphs, and $0.04\%$ were adults. The number of ticks [nymphs] per rodent was between $1$ [$0$] and $111$ [$15$] with a median number of ticks per rodent equals to $8$. Detailed data, on a yearly scale, are reported in Table~\ref{tab_capt}, while the fraction of nymphs observed in different year and grids is reported in Table~\ref{tab_capt_grid_year}. In Figure~\ref{topi_tempo_grid} the number of captured \textit{Apodemus flavicollis} per trapping session is shown for the whole nine year period and for different grids (from $A$ to $I$).

\subsection*{Data Analysis}\label{datanal}

\subsubsection*{Tick Burden Distribution}
Ticks patterns have usually been described as highly aggregated. Therefore, since the seminal works by Crofton \cite{Crofton1971}, Plowright et al. \cite{Plowright1977}, and those by Anderson and May \cite{AM1978} and \cite{MA1978}, the negative binomial (NB) probability distribution, 
\begin{equation}
q(k)={k+r-1 \choose k}(1-p)^{r}p^{k},\qquad\qquad k=0,1,2,\ldots
\end{equation}
has been considered suitable for describing macroparasite distribution on hosts. Here we used a maximum-likelihood-estimation (MLE) method to estimate the parameters $p$ and $r$ of the probability distribution of the tick burden on the entire dataset obtained by aggregating capture sessions and grids. In addition, we considered subsets of the original dataset composed by mice with large numbers of feeding ticks to evaluate the capability of the NB distribution to fit the tail of the parasite distribution. In particular, we estimated the parameters of the NB distribution on data characterised by $k\geq k_{\min}$, where $k_{\min}$ represents the threshold value of ticks per host above which the distribution is fitted. To evaluate the performance of the obtained fits we used Kolmogorov-Smirnov (KS) statistics. The goodness of each fit (GOF) was also evaluated through a bootstrap resampling procedure, generating $10^{3}$ synthetic data sets. The obtained $p$-value is defined as the relative number of times that the KS statistic of the fitted distributions on synthetic data exceeds that measured on real data. Therefore, the larger the $p$-value, the lower our confidence in rejecting the fit. We considered the conservative value $0.1$ as our threshold value, as suggested by Clauset and collaborators \cite{clauset2009}.

As a first alternative to the NB distribution, we considered a power-law probability distribution (PL), in its discrete version 
\begin{equation}
q(k)=Ak^{-\alpha},\qquad k=k_{\min},k_{\min}+1,k_{\min}+2,\ldots
\end{equation}
since it may represent a good candidate to describe the tail of the distribution \cite{Bisanzio2010}. We recall that $A^{-1}=\sum_{n=0}^{\infty}\frac{1}{\left(k_{\min}+n\right)^{\alpha}}$ represents the normalising factor of the probability distribution \cite{clauset2009}. To estimate the scaling parameter $\alpha$ of the distribution in such a way that the PL fits the data for $k\geq k_{\min}$, we followed the algorithm proposed by Clauset et al. \cite{clauset2009}. In short, the fitting procedure provides the best estimate for the parameters $k_{\min}$ (called $k_{\min}^{\textrm{PL}}$) and $\alpha$ by means of MLE and minimisation of KS statistics. Furthermore, bootstrap techniques were used to assess parameter standard deviations (std). We generated synthetic data and obtained a $p$-value through KS statistics to indicate the goodness of the fit, as for the NB distribution \cite{clauset2009}.

Another aggregated distribution used for describe pattern of macroparasites \cite{Elston2001} is the Poisson-LogNormal (PoiLN) distribution,
\begin{equation}
q(k)=\frac{\left(2\pi\sigma\right)^{-\frac{1}{2}}}{k!}\int_0^\infty \lambda^{k-1}e^{-\lambda}e^{\frac{-\left(\log(\lambda-\mu)^2\right)}{2\sigma}}\d \lambda, \qquad k=0,1,2,\ldots
\end{equation}
firstly introduced by Bulmer \cite{Bulmer1974} and used in several fields for its capability in describing aggregated data, e.g. \cite{Williams2012,Gonzales-Barron2011,Yin2005}. As for the NB distribution, we used a MLE method to estimate parameters $\mu$ and $\sigma$ on the entire data set. Uncertainty on the parameter estimation was assessed by bootstrap techniques. Moreover, in line with analysis performed for the NB distribution, we also explored the capability of the PoiLN distribution to describe a tick burden larger than a certain threshold $k_{\min}$ by coupling KS statistics and bootstrap procedures.

Finally, we compared the PL hypothesis in fitting the tail of the real data distribution with the two alternatives NB and PoiLN by a log-likelihood ratio (LLR) test for different values of $k_{\min}$. In particular, since the distribution models are non-nested, we used the method proposed by Vuong \cite{vuong1989} to understand whether the sign of such test was statistically significant or not.

Beyond the estimate of the ticks-per-host distribution, we also investigated how the tick burden distributions vary over time and whether a significant difference was observed when different time periods were considered. In particular, we investigated the tick aggregation patterns during periods characterised by low and high \textit{A. flavicollis} abundance. To achieve this goal we smoothed the time series of captured mice with a quadratic polynomial curve. %l'algoritmo sarebbe QR decomposizione e risoluzione
The parabola describing the mice abudance in a specific year and grid was normalised between $0$ and $1$ before isolating the time window where this normalised parabola was higher than a threshold value $\theta\in[0,1]$, thus identifying the peak time of mice abundance, as reported in Figure~\ref{time-window}. The distribution of ticks feeding on mice has been evaluated and compared considering in- and out-of-peak time periods for different values of $\theta$. We calculated the KS statistic between the in- (high abundance) and out-of- (low abundance) peak time distributions of tick burden, and we then compared the value observed in real data to a bootstrapped data set in order to establish whether this measure was statistically significant. For this purpose we generated $10^{5}$ synthetic in-and out-of- peak samples having the same size as the observed ones. As a test of soundness, we then calculated the fraction of the KS statistic that is larger in synthetic data than on real data.

\subsubsection*{Larval and Nymphal Aggregations Patterns on Mice}
A necessary condition for an effective non-systemic transmission of a pathogen is the coincidence of the larval and nymphal aggregation distributions on hosts \cite{Randolph1999}. Therefore, our first step was to examine the association between the number of larvae and that of nymphs on each host using Spearman's rank-correlation coefficient \cite{zar}. In particular, we preferred a non-parametric method rather than the more commonly used Pearson's correlation coefficient since tick distributions are aggregated (i.e. deviate from normal distribution) and we were more interested in any monotonic relations of our variables than in the linear relation depicted by the Pearson's coefficient. More in detail, a positive [negative] Spearman's coefficient would indicate that an increase in the number of nymphs per mouse is associated with an increase [decrease] of the number of larvae per mouse. Therefore, a positive Spearman's correlation coefficient could be interpreted as an indicator of the coincidence of the distributions, a zero coefficient could suggest the independence of the two distributions, and a negative coefficient, an uncommon result, would be an indicator of having two unimodal distributions with two asynchronous peaks. Moreover, to evaluate the significance of Spearman's coefficient (i.e. the probability that the same coefficient could be obtained by chance) we implemented a permutation test. In particular, we compared the evaluations on synthetic datasets with a reshuffled number of nymphs and on the original data and counted the number of times that the absolute value of Spearman's coefficient was larger than for the original data. The lower the sum, the higher our confidence in interpreting the association as significant.

To further evaluate the coincidence of tick stage distributions and the consequences on the non-systemic transmission of a pathogen, following Randolph et al. \cite{Randolph1999}, we evaluated the mean number of larvae cofeeding with a nymph on a host. In fact, the larger the mean, the larger the number of larvae that can potentially be infected via non-systemic transmission. After obtaining this empirical datum, we calculated the mean value for $10^3$ synthetic datasets where the number of nymphs was reshuffled, simulating independent distributions in order to have a more robust interpretation. After comparison of empirical and synthetic datasets, a significantly larger empirical mean number of larvae per nymphs gives evidences of coincident distributions, \cite{Randolph1999}.

\subsection*{Simulations of Tick-Borne Disease Spreading via Non-Systemic Transmission}\label{simu} 
In order to explore the impact of different parasite aggregation distributions on the spread of a TBEv-like pathogen where the main transmission route is through co-feeding, we performed extensive numerical simulations informed by the data about tick aggregation on mice. In this setting, tick larvae were not infective (transovaric transmission has been indicated as negligible \cite{Lindquist2008}), adults only rarely feed on mice (on our data set adults ticks are about $0.05\%$ of the total number of ticks feeding on mice), and the only transmission link that we considered was the co-feeding between infective nymphs and larvae. Therefore, the only actors in our model were nymphs and larvae feeding on hosts. Moreover, Ros\`a and collaborators suggested in a recent work devoted to the same geographical area \cite{Rosa2007b} that the larvae that feed in one year generally quest and feed as nymphs in the following year. Therefore, by adapting the Susceptible-Infected-Susceptible (SIS) model \cite{Anderson&May} to our purpose we assumed that nymphs are categorised as infective or not, that feeding larvae are susceptible and that some of them could eventually be infected by co-feeding with infective nymphs before moulting (thus becoming infective nymphs at time $t+1$). At each iteration $t$, with $t$ being a discrete number between $t_0$ and $t_{\max}$ and $\Delta t=1$ year, we assigned a number of ticks to each of the $N_h$ mice by drawing a sample from the considered distribution $q$. Then, on each mouse we said that of $k$ ticks feeding on it, $k f$ were nymphs and the other larvae (with $0<f<1$). These nymphs were larvae in the previous year and were possibly infected. Then, defining as $\pi_L(t-1)$ the prevalence among larvae after feeding at time $t-1$, we assumed that the prevalence at time $t$ among nymphs was $\pi_N(t)=\pi_L(t-1)$. Thus, the number of infective nymphs on a mouse that at time $t$ was parasitised by $k$ ticks was $k f \pi_N(t)$. Then, on each of the $N_h$ mice the co-feeding transmission between larvae and infective nymphs could occur with probability $\beta$ and we updated $\pi_L(t)$ accordingly to the fraction of larvae infected (i.e. the fraction of infective nymphs at next time step). The following meta-code summarises the epidemiological dynamic
\begin{enumerate}
	\item for $t$ between $t_0$ and $t_{\max}$:
	\begin{enumerate}
		\item for each mouse $i$, with $i$ between $1$ and $N_h$
			\begin{itemize}
				\item $k(i)$ is the number of ticks it feeds, being $k(i)$ a number drawn from the probability distribution $q$
				\item of the $k(i)$ ticks, $f k(i)$ are nymphs and the remaining larvae
				\item of the $f k(i)$ nymphs, a fraction $\pi_L(t-1) f k(i)$ are infective, the others are susceptible
				\item non-systemic transmission between infective nymphs and larvae on the same host occurs with probability $\beta$
			\end{itemize}
		\item $\pi_L(t)$ is updated as the fraction of larvae infected
		\item if $\pi_L(t)$ is equal to zero we stop the loop
	\end{enumerate}

\end{enumerate}
It is worth stressing that in the previous meta-code we did not consider ticks recovering from the infection, since we assumed that a feeding infective nymph at time $t$ will exit the infectious dynamics by moulting to the adult stage or dying.

We also modified the previous dynamics to deal with different distributions in tick aggregation as a function of seasonality. At each year $t$, we classified mice as observed during the mice peak activity (=$\gamma N_h$ mice, with $0<\gamma<1$) and observed out of the peak ($=(1-\gamma) N_h$). Therefore, we assigned the number of ticks feeding on mice according to the respectively aggregated distributions $q_{\textrm{IN}}$ and $q_{\textrm{OUT}}$. Moreover, since the larvae obtaining a blood meal at year $t$ will be nymphs at year $t+1$ without any other involvement in the epidemic spreading at year $t$, \cite{Rosa2007b}, these modifications to the meta-code are sufficient to suitably describe the seasonal variation in the epidemic process. More explicitly, the epidemic dynamic in the presence of seasonality in tick aggregation may be described by the following meta-code:
\begin{enumerate}
	\item for $t$ between $t_0$ and $t_{\max}$:
	\begin{enumerate}
		\item a fraction $\gamma$ of the $N_h$ mice are labelled as observed during mice peak activity (the remaining $=(1-\gamma) N_h$ as observed out of the peak window)
		\item for each mouse $i$, with $i$ between $1$ and $N_h$
			\begin{itemize}
				\item $k(i)$ is the number of ticks it feeds, being $k(i)$ a number drawn from the probability distribution $q_{\textrm{IN}}$, if the mouse was labelled as observed during the mice peak activity, or $q_{\textrm{OUT}}$, if not
				\item of the $k(i)$ ticks, $f k(i)$ are nymphs, the remaining larvae
				\item of the $f k(i)$ nymphs, a fraction $\pi_L(t-1) f k(i)$ are infective, the other susceptible
				\item non-systemic transmission between infective nymphs and larvae on the same host occurs with probability $\beta$
			\end{itemize}
		\item $\pi_L(t)$ is updated as the fraction of larvae infected
		\item if $\pi_L(t)$ is equal to zero we stop the loop
	\end{enumerate}
\end{enumerate}

%% ==================
%% ==================		

\section*{Results}

\subsection*{Ticks Burdens}
\begin{figure}[H]
\begin{center}
		\centering{\includegraphics{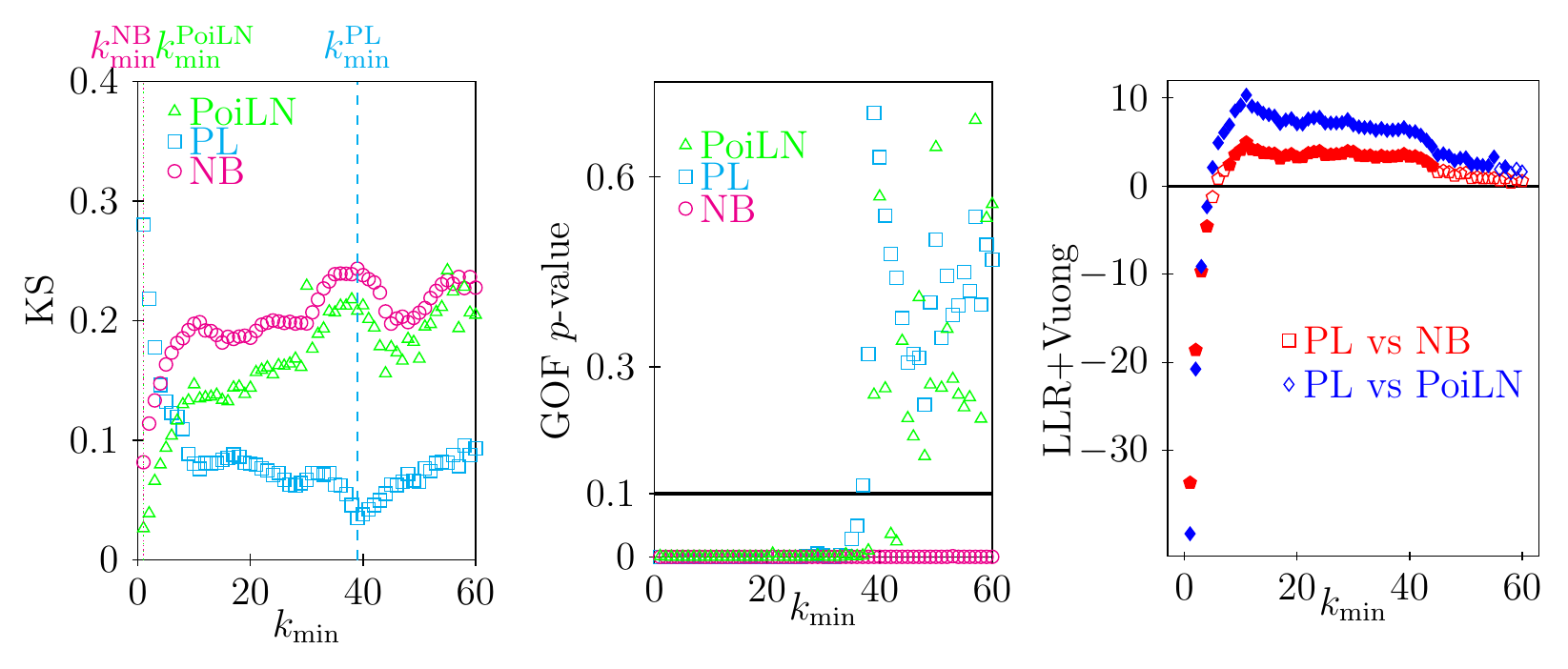}}
\end{center}
\caption{{\bf Comparison among fittings of distributions of ticks per host with different functions.} Left: Kolmogorov-Smirnov statistic between subsets of data above $k_{\min}$ and the fitting models on these subsets. Vertical dotted lines represent the optimum value of $k_{\min}$ for different models (NB: magenta; PoiLN: green; PL: cyan). For the NB and PoiLN models the optimum is observed for $k_{\min}^{\textrm{NB}}=k_{\min}^{\textrm{PoiLN}}=1$, i.e. on the entire data set, while for the PL model the optimum is reached for $k_{\min}^{\textrm{PL}}=38$. Center: goodness-of-fit $p$-value of fitting models on data larger than or equal to $k_{\min}$. As suggested by Clauset and collaborators \cite{clauset2009} for $p$-value greater than $0.1$ (horizontal line) the fitting model is a good description of the data. For NB the GOF is low ($p<10^{-3}$), suggesting the inappropriateness of the NB model in describing the data. The GOF of the PoiLN indicates that the model is appropriate only for large value of $k_{\min}$, thus simultaneously with large values of KS and therefore pointing out the low performance of the model. The PL fits should not be rejected for values of $k_{\min}$ larger than $35$ concurrently with the lowest value of KS. Right: Log-likelihood Ratio (LLR) test with Vuong's sign interpretation. Negative (positive) values suggest the alternative model NB (red) or PoiLN (blue) distributions are (are not) favoured in describing values larger than $k_{\min}$ when compared to PL. The horizontal line shows the sign threshold. Full marks show statistically significant tests ($p<0.05$)  while empty marks refer to non significant tests ($p>0.05$). %This picture points out that in a broad range of $k_{\min}$ the PL fitting should be preferred to the alternative models (PL is significantly preferred to NB for $k_{\min}\in[8,44]$ and to PoiLN for $k_{\min}\in[5,54]$).
} 
\label{gof_kmin} 
\end{figure}

The probability distribution of tick burden on mice was skewed and showed a heavy tail. The best fit of the NB distribution was obtained on the largest available subsets of data, i.e. with $k_{\min}=k_{\min}^{\textrm{NB}}=1$, see left panel of Figure~\ref{gof_kmin}. In this setting, the MLE method estimated $r=1.30$ ($95\%$ confidence intervals (CI) $=1.25,1.35$) and $p=0.10$ ($95\%$CI=$0.09,0.10$). However, the GOF of the NB distribution was very low $(p<10^{-3})$ for any value of $k_{\min}$, see central panel of Figure~\ref{gof_kmin}, thus giving evidence for rejecting the hypothesis of the NB functional form. Similarly, the best fit of PoiLN distribution was achieved on the largest subsets of data, ($k_{\min}=k_{\min}^{\textrm{PoiLN}}=1$, see left panel of Figure~\ref{gof_kmin}). In this case the estimated parameters were $\mu=1.96$ ($95\%$CI=$1.92, 1.99$) and $\sigma=0.99$ ($95\%$CI=$0.96, 1.02$). The GOF of the PoiLN, central panel of Figure~\ref{gof_kmin}, suggested that PoiLN was acceptable only for $k_{\min}>38$. However, for $k_{\min}>38$, the KS statistic displayed values that were too large to consider the PoiLN distribution appropriate for describing real data.

On the other hand, by fitting the tail of the distribution to a PL distribution, we found that the best fit was obtained for $k_{\min}=k_{\min}^{\textrm{PL}}=38$ (with a standard deviation of $5.83$), see left panel of Figure~\ref{gof_kmin}. This $k_{\min}$ value is matched with an estimated scaling parameter $\alpha=4.27$ (with standard deviation = $0.41$). The  GOF test ($p$-value larger than $0.1$) suggested that the optimum PL fit on the tail of the distribution should not be ruled out, and that the result holds for every PL fit with $k_{\min}>35$ see center panel of Figure~\ref{gof_kmin}. Finally, the LLR test highlighted that the PL fitting is to be preferred ($p<10^{-3}$) to the NB in describing the tail of the distribution for a large range of lower bounds, $k_{\min}\in[8,44]$, see right panel of Figure~\ref{gof_kmin}. Similarly, the PL is to be preferred to the PoiLN for $k_{\min}\in[5,54]$. Moreover, it is worth to stress that for values above $44$ ($55$) the sign of the LLR test still indicates the PL fit as the preferred one compared to the NB (and PoiLN), although the indication loses statistical significance due to the scarcity of available data.

\begin{figure}[!ht]
\begin{center}
		\centering{\includegraphics{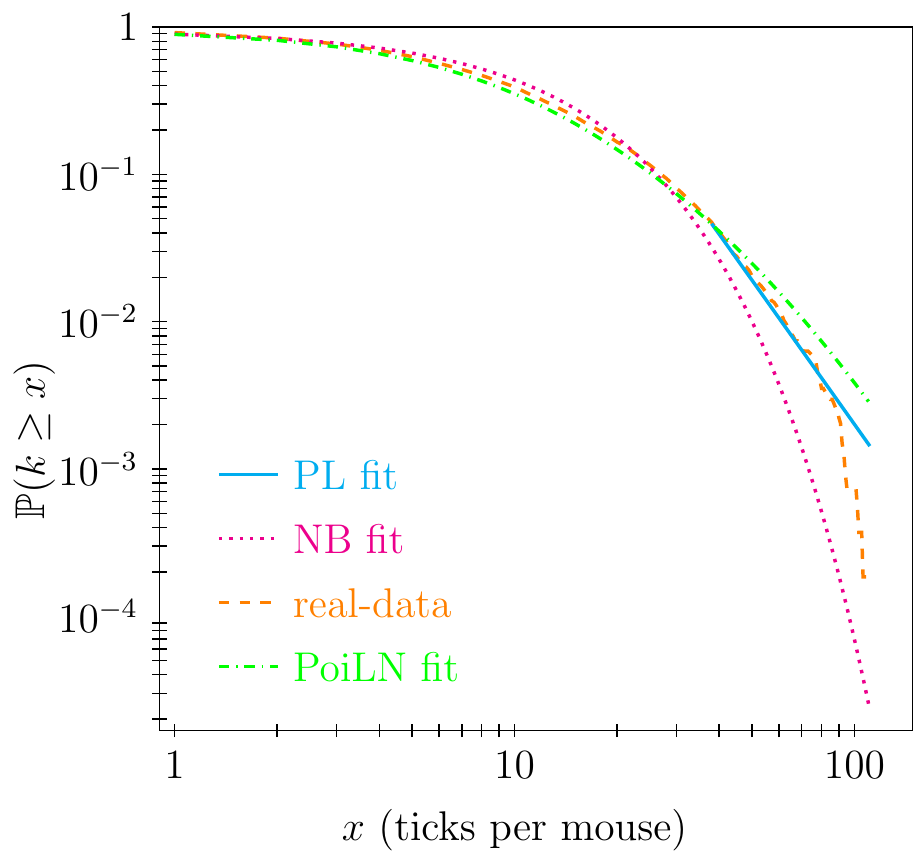}}
\end{center}
\caption{{\bf Complementary cumulative functions of number of ticks per host (real-data) with the best power-law (PL), negative binomial (NB),  and Poisson LogNormal (PoiLN) fit.} The PL fitting model shows high proximity to the tail of the real data distribution while the NB and the PoiLN fits appropriately describe the initial part of the distribution they describe the tail improperly.}
\label{distr} 
\end{figure}

In Figure~\ref{distr} we show the complementary cumulative probability distribution of the best fits resulting from $k_{\min}^{\textrm{NB}}=k_{\min}^{\textrm{PoiLN}}=1$ for NB and PoiLN distributions and $k_{\min}^{\textrm{PL}}=38$ for PL distribution against field data of the number of ticks per mouse. From this plot we noticed that above a certain number of ticks per mouse NB [PoiLN] under-estimates [over-estimates] the tail of the distribution (indeed both fits were statistically evaluated as very poor). At the same time, in agreement with statistical results summarised in Figure~\ref{gof_kmin}, we noticed that the PL fit in Figure~\ref{distr} more appropriately describes the right tail of the data distribution.

\begin{figure}[H]
\begin{center}
		\centering{\includegraphics{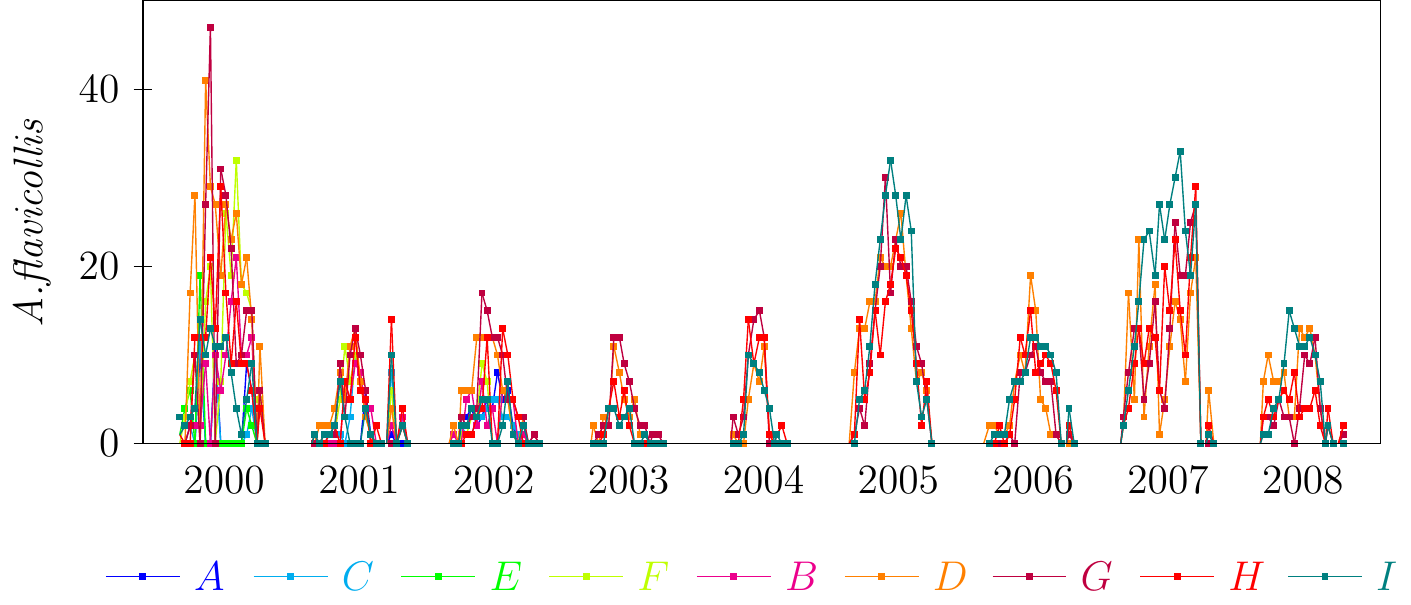}}
\end{center}
\caption{{\bf Temporal variation of \textit{A.flavicollis} mice abundance recorded in different grids} (labelled in different colours from A to I).}
\label{topi_tempo_grid} 
\end{figure}

\begin{figure}[H]
\begin{center}
		\centering{\includegraphics{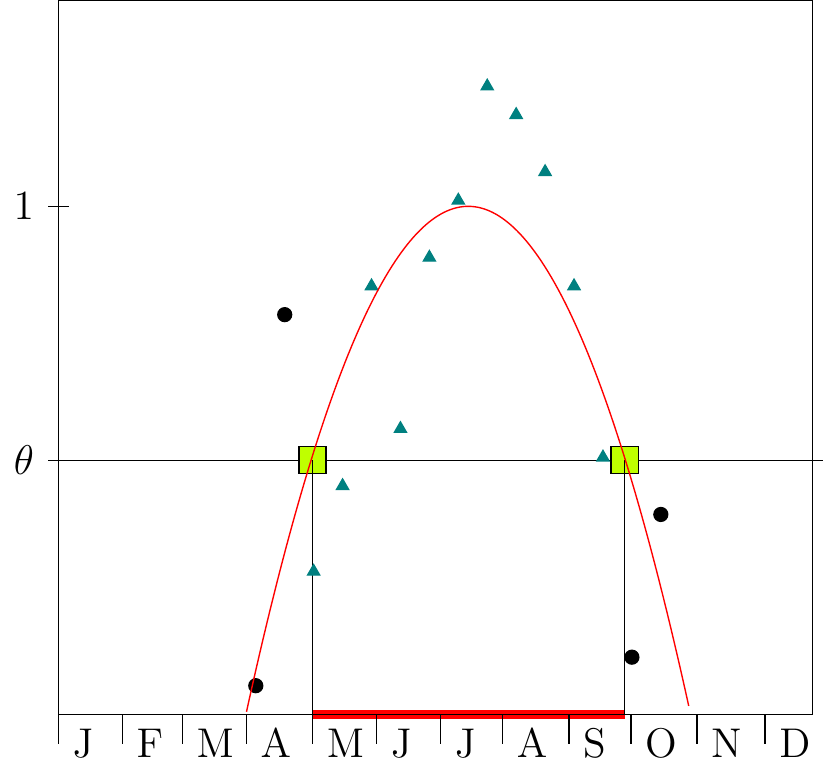}}
\end{center}
\caption{{\bf Detection of seasonal abundance time-windows.} The time series of captured mice has been interpolated by a quadratic polynomial curve. By normalising the obtained parabola to unity and setting a threshold $\theta$ ($=0.5$ in the example), we identify mice captured in high abundance season, those above the threshold $\theta$ (triangles), and mice captured in low abundance period, those below the threshold (circles).}
\label{time-window}
\end{figure}

The number of mice captured in different years and grids showed strong seasonal patterns as reported in Figure~\ref{topi_tempo_grid}. For each grid and each year we defined two separate periods depending on the mice abundance as defined in section ``Data Analysis'' and sketched in Figure~\ref{time-window}. Imposing a threshold $\theta$, for each year and grid we identified a time window of high mice abundance. With $\theta=0.5$ we found significant evidence that the distribution of ticks on mice within the abundance peak was different from that observed outside. Indeed, the fraction of the KS measures calculated on the synthetic samples lower than the real-data KS statistic was almost $98\%$, thus indicating very low confidence in obtaining the same measurement by chance. The same statistical evidence was also obtained by using different time window thresholds (such as, $\theta=0.4$ and $0.6$).

\begin{figure}[H]
\begin{center}
		\centering{\includegraphics{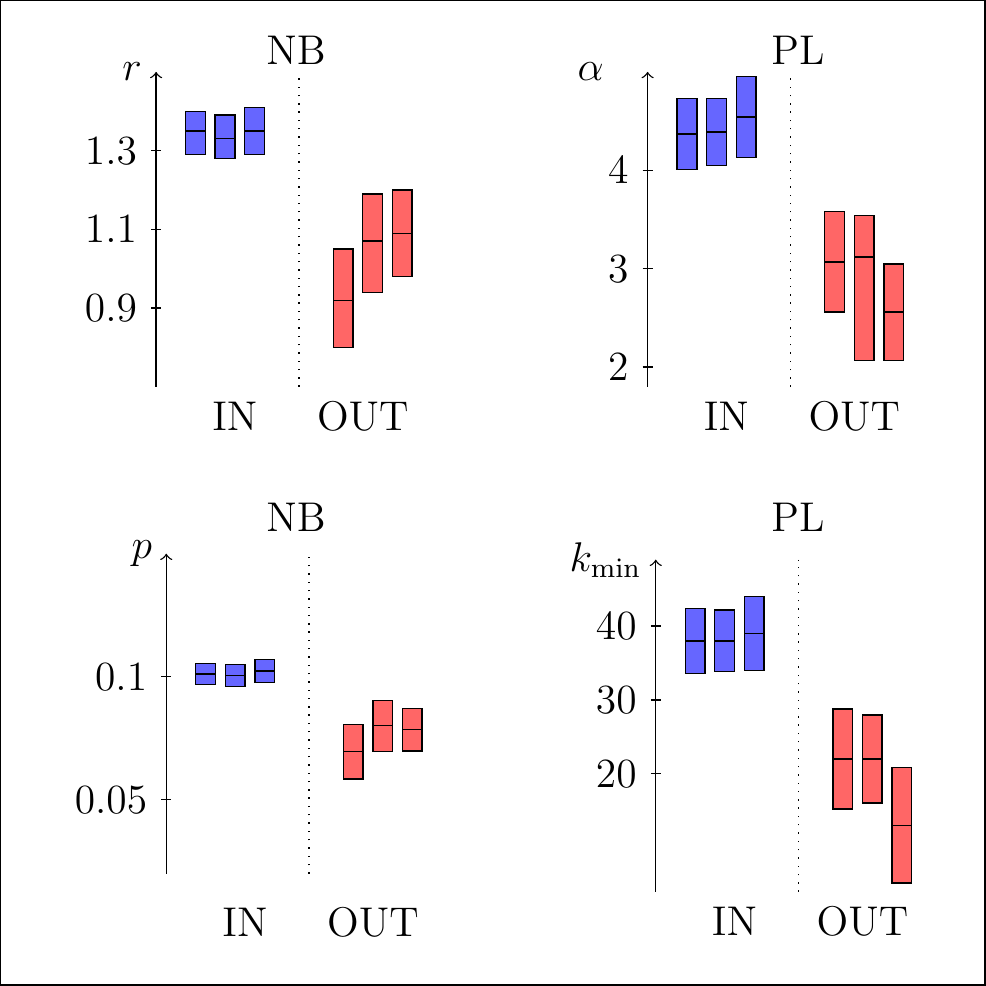}}
\end{center}
\caption{{\bf Estimated parameters of different distributions (NB on left and PL on right) obtained inside (blue) and outside (red) of the mice peak abundance time window.} Time windows are defined by $\theta=0.4,0.5,0.6$ (from left to right for each subsets). Vertical bars indicate best model fits (central horizontal lines) with their uncertainties that are $95\%$ confidence interval for NB models while standard deviations for PL models.}
\label{tw} 
\end{figure}

On the data sets classified as inside (IN) and outside (OUT) the time window of mice abundance peak, we fitted for different time-window lengths ($\theta=0.4,0.5,0.6$) the parameters $r$ and $p$ for NB distribution (Figure~\ref{tw}, left panels) and $\alpha$ and $k_{\min}$ for PL distribution (Figure~\ref{tw}, right panels). We observed a larger PL scaling parameter $\alpha$ inside the mice abundance peak than outside (two-sample $t$-test output: for $\theta=0.5$ $t$-statistic=$-74.95$,  df$=1931$, $p<10^{-3}$) indicating a larger heterogeneity in tick burden outside the abundance peak time. Moreover the GOF test indicated a rejection of the NB fit in both sets (IN and OUT) with $\theta=0.5$. On the other hand, the GOF test with $\theta=0.5$ showed that the PL model cannot be ruled out in both sets ($p$-value $>0.1$) and the LLR test indicated that the PL fitting outperforms the NB model ($p$-value $<0.05$) in the estimates both inside and outside the peak time window. 

The distribution of larvae and nymphs on mice are coincident rather than independent, and indeed the same $20\%$ most infested hosts feed both $55\%$ of the nymphs and $54\%$ of the larvae. Moreover, Spearman's correlation coefficient measured on the number of larvae and nymphs on mice was positive ($0.24$) and the probability that this coefficient was detected by chance was very low (the empirical value was the largest if compared to those evaluated in $10^3$ reshuffled samples). In addition, the mean number of larvae co-feeding with a nymph is about $23$ which is almost double the value that would be seen if the distributions were independent (mean equal to $12$).

\subsection*{Non-systemic disease spreading simulations}
\label{simulatio} 

To start, we simulated the non-systemic disease spreading of a TBE-like pathogen with a fraction $f$ of nymphs among ticks equals to $2\%$, close to the one observed in our real data (cfr. Table~\ref{tab_capt_grid_year}), $5\%$, and $10\%$, as in literature \cite{Randolph1995,Randolph1999}. We consider the empirical distribution observed on the entire data set. We fixed the number of hosts to $N_h=10^4$ which, together with the considered distribution, resulted in a number of vectors pairs equal to $N_V\sim10^5$. In our simulations, we explored the effects of $\beta$, the infection probability, on the observed prevalence at the final time step, $\pi_L(t_{\max})$, with $t_{\max}=1000$. (We observed that $t_{\max}=1000$ was larger enough to allow the prevalence to converge toward an endemic pseudo-equilibrium or the disease-free equilibrium). For each $\beta$ we allowed $200$ simulations to run starting from an initial prevalence of $\pi_N(t_0)=1\%$. In Figure~\ref{prev_prima} we plotted the prevalences (median value, interquartile intervals and the $95\%$CI) observed at equilibrium as a function of the transmission probabilities, $\beta$. Results showed that the larger the fraction of nymphs among ticks feeding on mice, the larger the probability of pathogen invasion and the infection prevalence.

Then, we explored the effects of different tick burden distributions on the spread of infection. To this end we considered four distributions: PL, NB, PoiLN, and the empirical distribution on the entire data set (aggregated on capture sessions and grids). For synthetic distributions we considered the actual observed distribution below the estimated $k_{\min}$, while we used the best fit of synthetic distributions to describe values greater than $k_{\min}$. Again, we fixed the number of hosts to $N_h=10^4$. It is worth stressing that in the synthetic samples generated from these distributions we observed some features similar to those observed in real-data. For instance, the number of nymphs was positively associated with that of larvae and more particularly a nymph co-fed with a mean number of larvae similar to that observed in reality (for PL the mean number was $23$, for NB $20$, and for PoiLN $27$).

Results, plotted in Figure~\ref{prev} for $f=2\%$ and in S2-text for $f=5\%$ and $f=10\%$, corroborated the hypothesis that the transmission probability needed for the pathogen to become endemic is driven by the shape of the tail of the distributions. In particular, we noticed that for the PoiLN distribution (the one with larger fitted tail) the epidemic threshold is the lowest, while for the NB distribution (the one with smaller fitted tail) the infection probability needed for invasion is the highest. Not surprisingly, the PL, which has the best performances in fitting the tail of the empirical distribution, is the one for which the prevalences at equilibria better resemble those observed in simulations using the empirical distribution. We also performed some sensitivity analysis on parameter distributions, further highlighting that the larger the tail of the distribution, the lower the epidemic threshold (see \ref{SM1}). In addition, sensitivity analysis on the fraction of nymphs ($f$) showed that $f$ does not qualitatively influence the epidemic behaviour (see \ref{SM2}).

Furthermore, we investigated the effect of differences in the distribution of the tick burden as a function of the abundance of mice on the spreading of a non-systemic infectious disease. To this end, we fixed $\gamma=0.89$, as measured in the dataset, and as $q_{\textrm{IN}}$ we considered a PL with exponent $\alpha_{\textrm{IN}}=4.39$ as estimated with $\theta=0.5$. In a similar way, we assumed as $q_{\textrm{OUT}}$ a PL distribution with exponent $\alpha_{\textrm{IN}}=3.48$. For both $q_{\textrm{IN}}$ and $q_{\textrm{OUT}}$ we further set $k_{\min}=5,10,15$. Results are summarised in Figure~\ref{prev_tw_exa}, from which it could be inferred that the epidemic outcome was strongly influenced by the different distributions of feeding ticks according to mice abundance. We consistently observed that the transmission probability needed for the pathogen to effectively spread was smaller when the time windows identified by mice abundance are considered.

%% ==================
%% ==================

\section*{Discussion}

Tick aggregation on hosts is the result of several complex interactions of biotic and abiotic factors, such as host exposure and susceptibility to ticks, ticks' phenology and host behaviour, environmental factors, availability of resources, and others \cite{Wilson2002,Brunner2008}. Historically, the NB distribution has been preferred to the Poisson distribution to describe parasite heterogeneity across hosts because it suitably reproduces overdispersed observations. It has also been widely used in empirical \cite{Randolph1975,Shaw1995,Shaw1998,Kiffner2011} and theoretical studies \cite{Rosa2003,Rosa2007,Harrison2012}. However, \textit{fat tailed} distributions other than the NB one can also adequately reproduce tick aggregation, as shown by Elston et al. \cite{Elston2001} and Bisanzio and collaborators \cite{Bisanzio2010}.

\begin{figure}[H]
\begin{center}
		\centering{\includegraphics{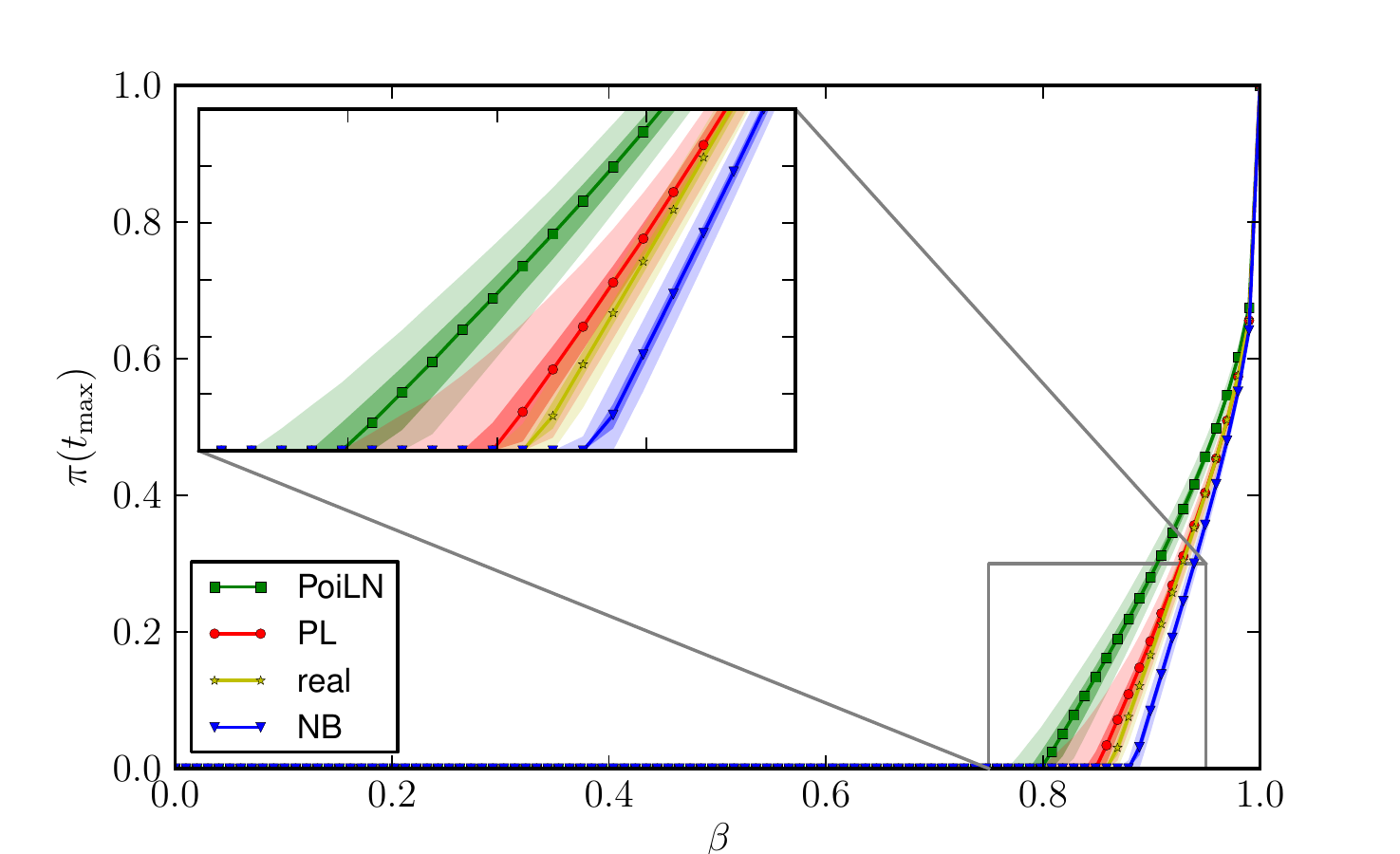}}
\end{center}
\caption{{\bf Median (line), interquartile (darker area) and $95\%$ confidence intervals (lighter area) of the final prevalence as a function of the transmission probability,} for different fitting distributions (PL, NB and PoiLN). Other parameters are $N_h=10^4$, $t_{\max}=10^3$, and $f=2\%$.}
\label{prev} 
\end{figure}

Through the use of an extensive data set of feeding \textit{Ixodes ricinus} ticks on mice, we showed that a PL distribution is better able to describe the right tail of the tick distribution on hosts than a NB or a PoiLN distribution (see Figure~\ref{gof_kmin} and \ref{distr}). This finding may have relevant epidemiological consequences, since it is well documented that the heterogeneity of contact distributions among individuals has large impacts on pathogen spread and persistence \cite{Becker1990,Ball1997,lloyd2001,Pastor2001,newman2002,May2001,Barrat2008}. In fact, it has been demonstrated \cite{pastor_rapid_2002} that the minimum transmission probability for a pathogen to spread on a network, the so-called epidemic threshold, is driven by the first and the second moment of this distribution. In particular, Pastor-Satorras et al. \cite{pastor_rapid_2002} demonstrated that the larger the heterogeneity, the lower the epidemic threshold for the pathogen to spread, with an interesting behaviour in infinite size network showing a zero epidemic threshold \cite{Pastor2001}. Thus, the epidemiological inferences on the spread of a pathogen are highly influenced by the characterisation of the connectivity distribution and in particular by the distribution tail (i.e. the heterogeneity). Our results corroborate those findings and generalise them in a different framework and for more complex transmission routes, i.e. a vector-host network for non-systemically transmitted diseases. In particular, we found that the tail of the distribution of the number of ticks per rodent highly influences pathogen spreading (see Figure~\ref{prev} and S1-text). Furthermore, it is worth remarking that although the tail of the distribution as defined here represents about $5\%$ of the entire data set, our simulation findings suggest that this small part of the distribution is crucial for pathogen invasion.

\begin{figure}[H]
\begin{center}
		\centering{\includegraphics{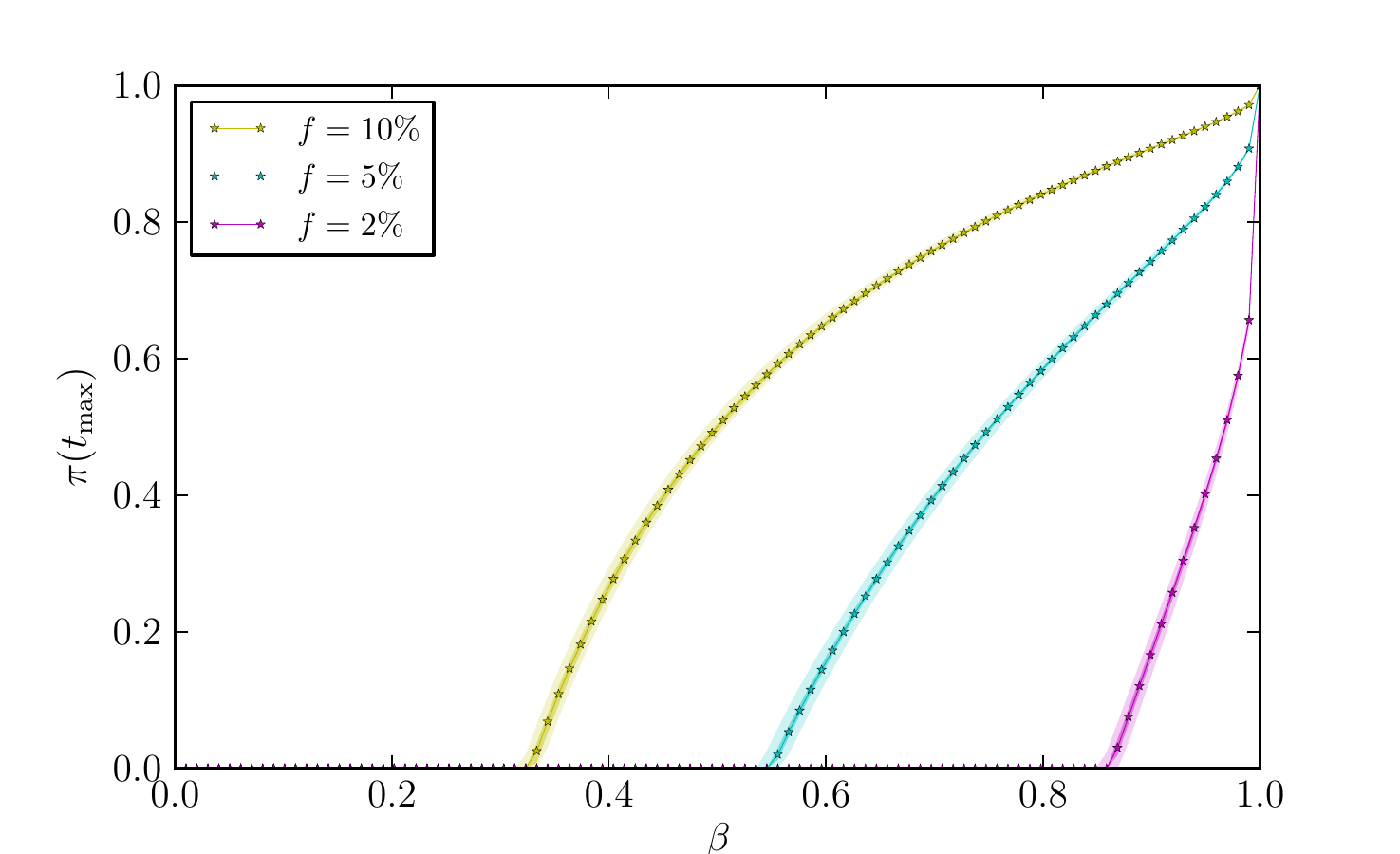}}
\end{center}
\caption{{\bf Median (line), interquartile (darker area) and $95\%$ confidence intervals (lighter area) of the final prevalence as a function of the transmission probability,} for different values of $f(=2\%,5\%,10\%$), fraction of nymphs among ticks on a mouse, and by describing the ticks aggregation with the empirical distribution. Other parameters are $N_h=10^4$, $t_{\max}=10^3$.}
\label{prev_prima} 
\end{figure}

We also confirm that the probability of pathogen invasion and the infection prevalence are strongly influenced by the fraction $f$ of nymphs on the total feeding ticks on mice (Figure~\ref{prev_prima} and S2-text). The co-occurrence of larvae and nymphs on competent hosts is in fact essential for the horizontal transmission of non-systemic transmitted tick-borne pathogens, such as TBE, and it has been documented, both empirically and theoretically, that it could be a key factor in creating TBE hotspots, \cite{Cagnacci2012,Bolzoni2012}.

Our conclusions confirm previous findings showing that the distribution of ticks on rodents may significantly affect the spread of infections \cite{Brunner2008,Bisanzio2010,Calabrese2011}, especially for non-viraemic transmitted diseases such as TBE \cite{Rosa2003,Perkins2003,Harrison2012}. Under the hypothesis of a NB distribution of ticks across hosts, both Ros\`a et al. \cite{Rosa2003} and Harrison and collaborators \cite{Harrison2012} showed that highly coincident and aggregated distributions favour the establishment of TBEv. However, highly heterogeneous degree distributions do not necessary imply a higher spread of disease. Indeed, Piccardi et al. \cite{Piccardi2008} showed that scale-free networks can be much less efficient than homogeneous networks in favouring the disease spread in the case of a nonlinear force of infection.

The correct description of tick aggregation on hosts could dramatically affect disease control strategies: for instance, Perkins \cite{Perkins2003} emphasised that an optimised control effort targeted on highly parasitised mice, also identified as sexually mature males of high body mass, could significantly lower the transmission potential. On the other hand, Brunner and colleagues \cite{Brunner2008} observed that the identification of individuals which fed a disproportionate number of ticks (and that can therefore act as superspreaders) can be challenging, since simple covariates such as sex, age or mass do not entirely explain the differences in parasite burden.

\begin{figure}[H]
\begin{center}
		\centering{\includegraphics{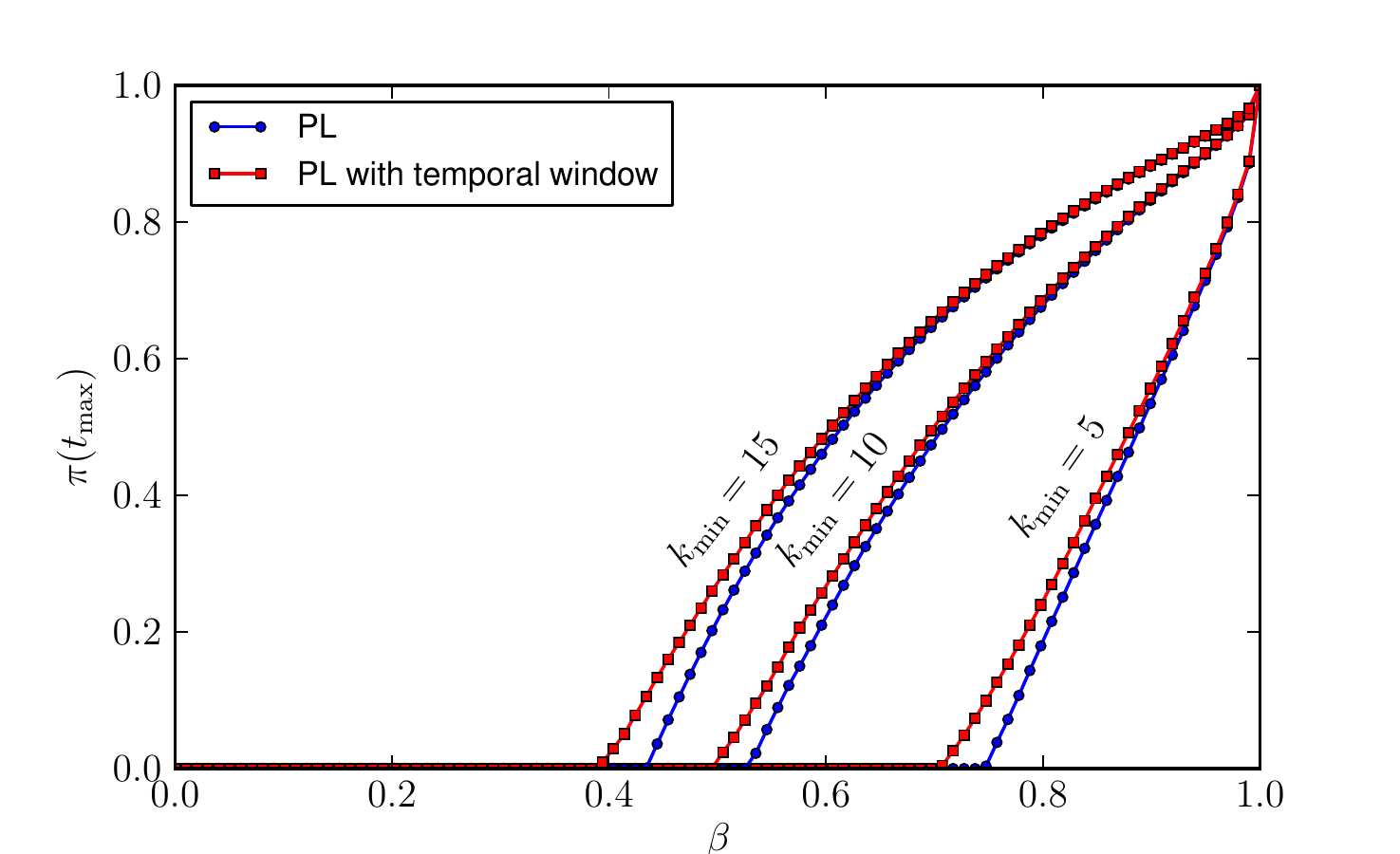}}
\end{center}
\caption{{\bf Median of the final prevalence as a function of the transmission probability.} A PL distribution of vectors-per-host has been considered in all scenarios. Simulations that consider different aggregation behaviours according to the temporal window of mice abundance (red) are compared with others with a fixed distribution (blue). Other parameters are $N_h=10^4$, $t_{\max}=10^3$, and $f=10\%$.}
\label{prev_tw_exa} 
\end{figure}

In order to fully understand the different tick attachment behaviours on hosts, we identified different time windows related to rodent seasonal dynamics. Using this approach we found that the distribution of ticks on mice may vary across the season, with higher aggregation heterogeneity in periods of low rodent abundance and lower aggregation heterogeneity during the peak of host abundance (see Figure~\ref{tw}). We also showed that seasonal aggregation patterns, characterised by larger \textit{tails} in time periods of low host abundance, enhance the spread of non-viraemic transmitted diseases (see Figure~\ref{prev_tw_exa}). Shaw and collaborators \cite{Shaw1998} observed significant variations in the degree of aggregation between host subsets -- stratified by sex, age, space or time of sampling -- in several host-parasite systems. In agreement with our results (lower aggregation in period of high mice abundances as shown by estimated exponents of PL), they found that aggregation in copepod (\textit{Lepeophtheirus pectoralis}) infesting plaice (\textit{Pleuronectus platessa}) decreases during summer months. They mainly ascribed the observed variation to significant differences in mean parasite burden among months. On the other hand, we did not find significant differences in tick burden inside and outside the window of high rodent abundance. Specifically, in the case of $\theta=0.4,0.5,0.6$, the average number of ticks per host were $11.96,11.96,11.84$ inside the window of high rodent abundance and $12.34,12.23,12.78$ outside and the differences between inside and outside are not statistically significant (permutation tests, $p>0.05$). However, the second moment of the number of ticks per host drastically changed between high and low abundance periods, driving the difference in the aggregation distributions observed in the two time windows. Seasonal variations in resource availability and host abundance can have a significant effect on the space used by mice. Males and females tend to respond to these changes in different ways, since space use for females is driven largely by food availability, whereas the distribution of males is related primarily to mating opportunities. Yellow-necked mouse (\textit{A. flavicollis}) females exhibited reduced spatial exclusivity and larger home ranges during lower food availability while males varied their spatial distribution accordingly by also expanding their home ranges \cite{Stradiotto2009}. An inverse relationship between population density and home range sizes has also been observed in wood mice (\textit{Apodemus sylvaticus}) \cite{Wilson1993}. Consequently, in periods of low rodent abundance more mobile rodents, especially males, are more likely to hit a patch of larval ticks. As a result, these individuals would harbour a large amount of ticks and increase the aggregation of tick distribution among the rodent population. On the other hand, tick density is usually lower in periods of low rodent abundance, and the average tick burden would decrease for the rest of the population, especially females, balancing the overall tick burden. On the contrary, during times of high abundance mice move less and ticks would be distributed more evenly among the rodent population resulting in the observation of a lower aggregation in tick distribution during the peak of rodent
abundance.

Our primary goal was to help understand the role of tick aggregation across mice on the spread of non-viraemic transmitted diseases through a simple and general transmission model. Other works -- such as \cite{Rosa2003,Norman2004,Rosa2007,Bolzoni2012} -- described in very fine detail the transmission of vector-borne diseases, introducing different transmission routes, tick stages and alternative hosts in the epidemic model. For instance, Norman and colleagues \cite{Norman2004} demonstrated through an epidemiological model that non-viraemic transmission could have non-negligible effects on the persistence of a disease like the Louping ill. Here, considering the non-systemic transmission only, we explored the effect of using different theoretical functional forms to describe the tick burden on hosts. By estimating parameters of the burden distributions on a very detailed data set, we defined a simple and transparent transmission model that explicitly takes into account the real contact pattern of vectors and hosts in the description of a non-systematically transmitted vector-borne disease. In this way we were able to emphasise that, while the NB and PoiLN models can sufficiently fit the whole real distribution, the PL model represents a better fit for the distribution tail. Furthermore, the vector perspective approach used in our model gives better insights into the dynamics of non-systemic transmitted pathogens respect to host perspective models that were more commonly and widely used in this context \cite{Rosa2003,Rosa2007,Norman2004,Bolzoni2012}. In addition, epidemiological simulations parameterised by the fitted tick burden distributions highlighted the epidemiological consequences of describing tick aggregation on hosts trough distributions with different tails, showing that the shape of the tail distribution has a non-negligible influence on pathogen persistence. Future works will be devoted to extend the present findings to more complex transmission dynamics (e.g. including viraemic or transovaric transmission), in order to assess the effect of a PL decay of the distribution for a wider range of vector-borne diseases.

\section*{Acknowledgements}

Authors wish to thank all the field assistants from the Department of Biodiversity and Molecular Ecology at Fondazione Edmund Mach. Authors wish to thank Bryan N. Iotti for the painstaking proofreading of the whole manuscript.

\section*{Financial Disclosure}
This study was partially funded by European Union (EU) Grant FP7-261504 \href{http://www.edenext.eu}{EDENext} and is catalogued by the EDENext Steering Committee as EDENext265. The contents of this publication are the sole responsibility of the authors and do not necessarily reflect the views of the European Commission. LF acknowledges support from the \href{http://www.progettolagrange.it/en/}{Lagrange Project}, CRT and ISI Foundation.

\bibliography{biblio}

\newpage
%\section*{Figure Legends}
\section*{Figures}



\section*{Tables}

\begin{table}[H]

\centering
\caption{\textbf{Basic descriptive statistics for empirical data.}}
\scalebox{0.75}{
				\begin{tabular}{ l|cccccccccc}
				&$2000$&$2001$&$2002$&$2003$&$2004$&$2005$&$2006$&$2007$&$2008$\\\hline
				\# of grids&$9$&$8$&$8$&$4$&$4$&$4$&$4$&$4$&$4$\\
				\# of sessions&$16$&$15$&$15$&$13$&$10$&$15$&$15$&$16$&$14$\\
				\# of mice&$1207$&$356$&$434$&$137$&$187$&$854$&$327$&$897$&$323$\\
				sum of feeding ticks&$14376$&$7073$&$6550$&$2426$&$3063$&$7361$&$4077$&$5821$&$4685$\\
				median of ticks per rodent&$9$ &$14$ &$11$ &$14$ &$11$ &$6$ &$8$&$4$ &$10$\\
				ranges of ticks per rodent&$(1,103)$&$(1,102)$&$(1,78)$&$(3,88)$&$(1,95)$&$(1,111)$&$(1,93)$&$(1,85)$&$(1,77)$\\
				nymphs fraction&$0.5\%$&$2.2\%$&$2.7\%$&$1.1\%$&$3.0\%$&$0.8\%$&$2.0\%$&$0.7\%$&$1.0\%$\\
				\end{tabular}}
\begin{flushleft}Number of trapping grids, trapping sessions, total number of \textit{A. flavicollis} captures for different years, sum of feeding ticks, median and ranges of the number of ticks per rodent, and mean number of nymphs fraction among feeding ticks. \end{flushleft}
\label{tab_capt} 
\end{table}

\begin{table}[H]
\centering
\caption{\textbf{Nymphs to total ticks ratio for observed feeding ticks on mice.}}
				\begin{tabular}{ l|ccccccccc}
				&$2000$&$2001$&$2002$&$2003$&$2004$&$2005$&$2006$&$2007$&$2008$\\\hline
				$A$&$1.1$&$1.0$&$4.0$&$0.3$&$2.0$&$1.1$&$1.8$&$0.6$&$1.6$\\
				$B$&$0.5$&$1.6$&$0.7$&$1.0$&$4.6$&$0.6$&$1.8$&$0.7$&$0.1$\\
				$C$&$0.4$&$1.2$&$2.8$&$1.3$&$1.7$&$0.4$&$3.1$&$1.2$&$1.4$\\
				$D$&$0.2$&$1.8$&$3.6$&$1.0$&$3.9$&$0.9$&$1.4$&$0.5$&$0.5$\\
				$E$&$0.4$&$4.4$&$1.6$&-&-&-&-&-&-\\
				$F$&$0.9$&$2.8$&$2.2$&-&-&-&-&-&-\\
				$G$&$0.3$&-&-&-&-&-&-&-&-\\
				$H$&$1.1$&$1.5$&$3.0$&-&-&-&-&-&-\\
				$I$&$0.7$&$10.4$&$3.3$&-&-&-&-&-&-\\
				\end{tabular}
\begin{flushleft} Percentage of feeding nymphs on the total feeding ticks observed on mice in different years ($2000$-$2008$) and grids ($A$-$I$). \end{flushleft}
\label{tab_capt_grid_year} 
\end{table}

\appendix
\renewcommand{\thesection}{SM.\arabic{section}}
\section{Sensitivity Analysis of Distribution Parameters on Epidemic Spreading}
\label{SM1}
In this section we explored the effect of parameters of tick burden distributions not at the best fit on the epidemic spreading. Results suggested that the larger the heterogeneity caused by parameters, the lower the epidemic threshold.

\begin{figure}[H]
\begin{center}
		\centering{\includegraphics{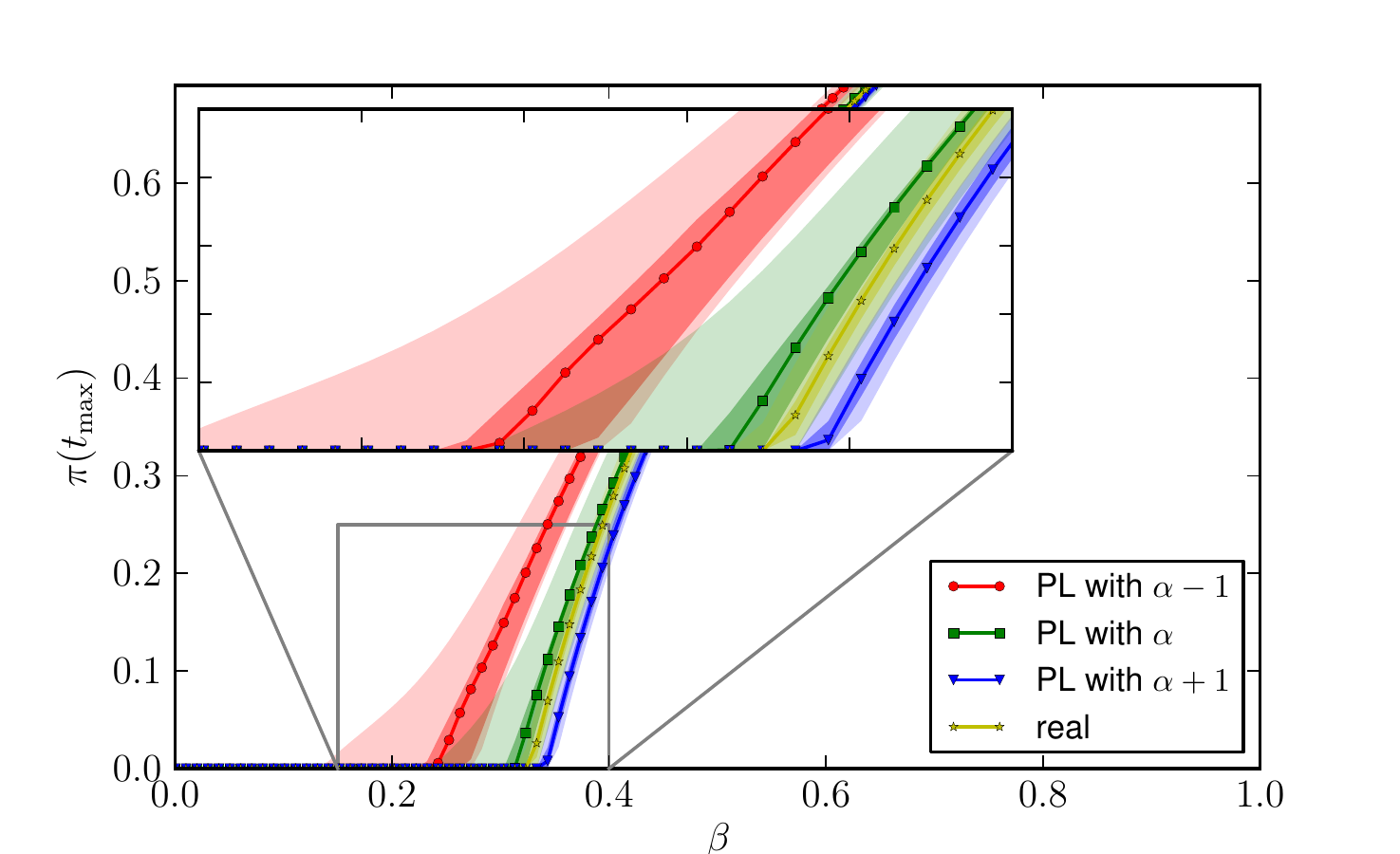}}
\end{center}
\caption{{\bf Median (lines), interquartile (darker areas) and $95\%$ confidence intervals (lighter areas) of the final prevalence as a function of the transmission probability.} Ticks burdens are described by PL distribution with different exponents (see legend). In particular, we explore the sensitivity of this function to variations in $\alpha$, which represents the best fit parameter on the empirical data. The long-term prevalence obtained with tick burdens sampled from the empirical distribution is also plotted as benchmark.}
\label{SM1_1} 
\end{figure}
\begin{figure}[H]
\begin{center}
		\centering{\includegraphics{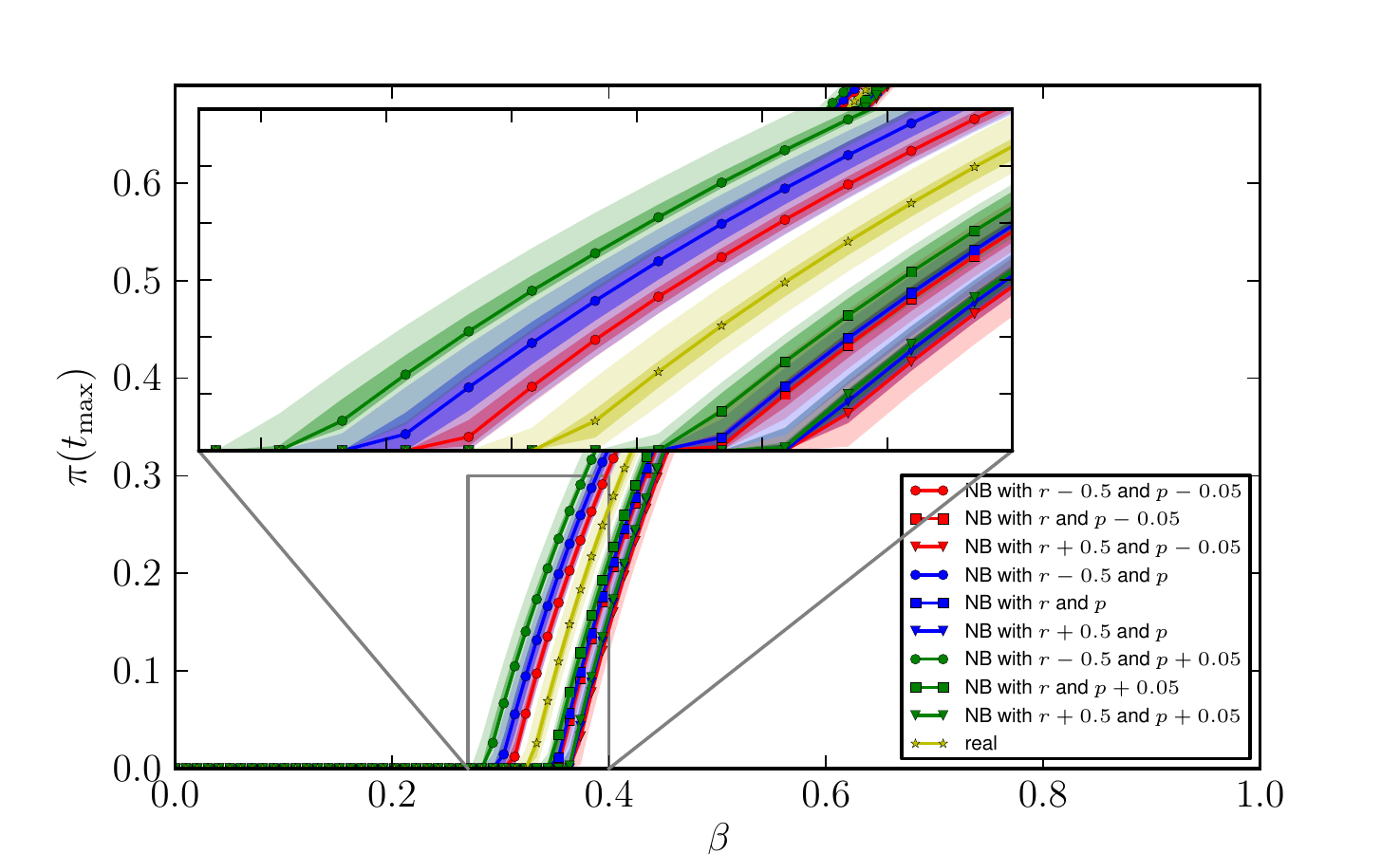}}
\end{center}
\caption{{\bf Median (line), interquartile (darker areas) and $95\%$ confidence intervals (lighter areas) of the final prevalence as a function of the transmission probability.} Ticks burdens are described by NB distribution with different parameters (see legend). In particular, we explore the sensitivity of this function to variations in $r$ and $p$, which represent the best fit parameters on the empirical data. The final prevalence obtained with tick burdens sampled from the empirical distribution is also plotted as benchmark.}
\label{SM1_2} 
\end{figure}
\begin{figure}[H]
\begin{center}
		\centering{\includegraphics{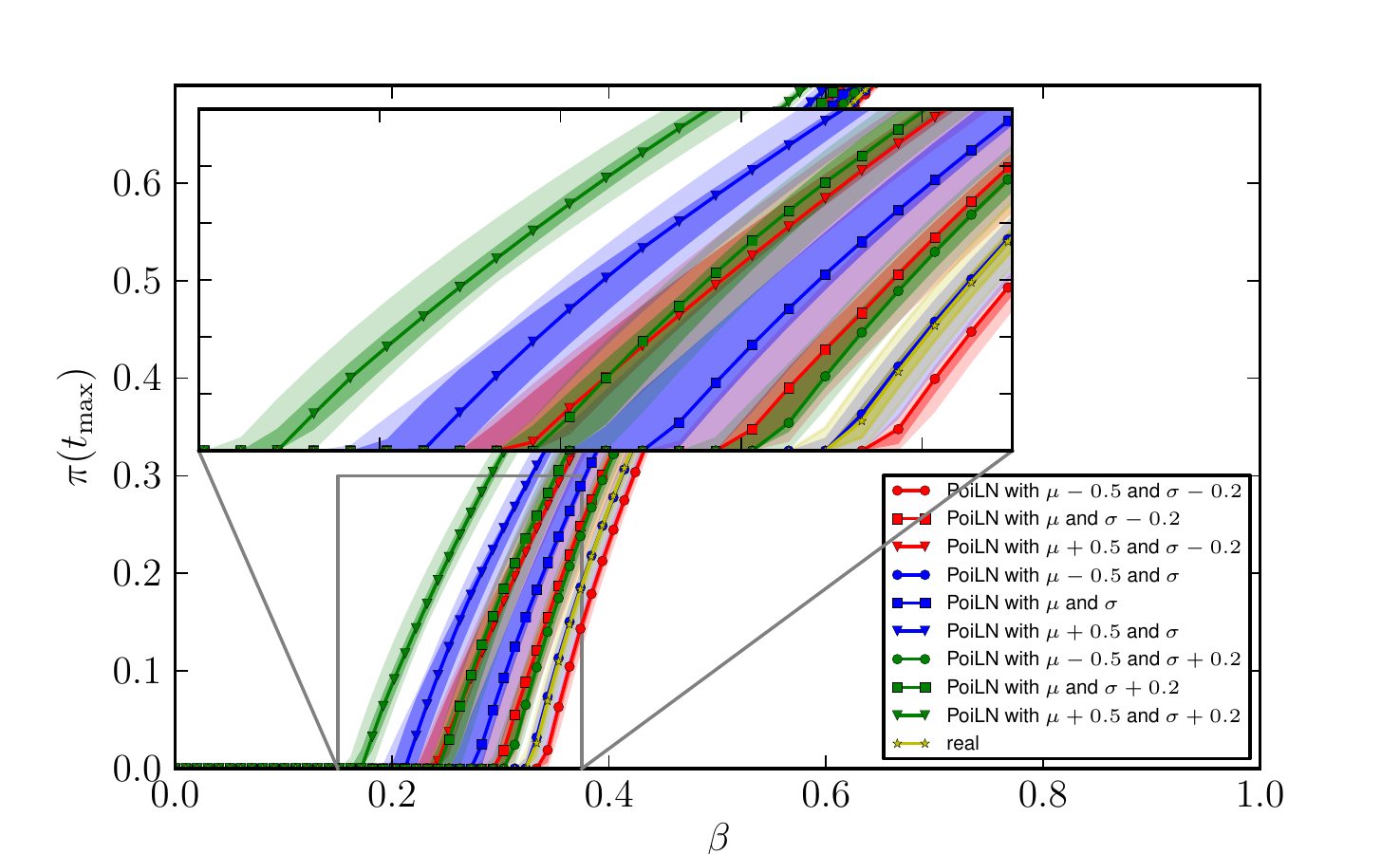}}
\end{center}
\caption{{\bf Median (line), interquartile (darker areas) and $95\%$ confidence intervals (lighter areas) of the final prevalence as a function of the transmission probability.} Ticks burdens are described by PoiLN distribution with different couple of parameters. In particular, we explore the sensitivity of this curve to variations of $\mu$ and $\sigma$, the best fit parameters on the empirical data. The final prevalence obtained with tick burdens sampled from the empirical distribution is also plotted as benchmark.}
\label{SM1_3} 
\end{figure}
\newpage
\section{Sensitivity Analysis of $f$, Fraction of Nymphs among Ticks}
\label{SM2}
\begin{figure}[H]
\begin{center}
		\centering{\includegraphics{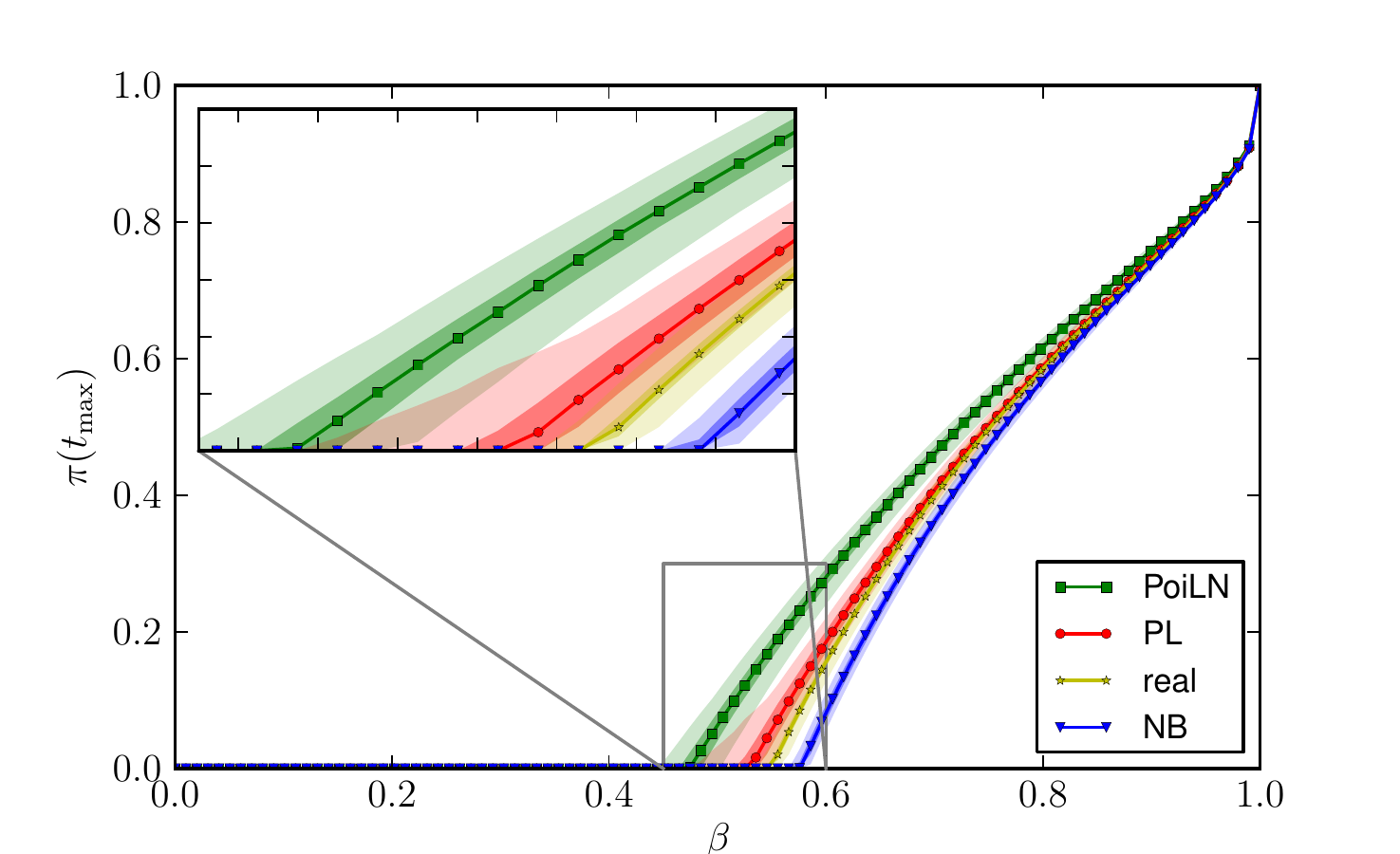}}
\end{center}
\caption{{\bf Median (lines), interquartile (darker areas) and $95\%$ confidence intervals (lighter areas) of the final prevalence as a function of the transmission probability.} $f$, fraction of nymphs among ticks is fixed to $5\%$.}
\label{prev_per5} 
\end{figure}

\begin{figure}[H]
\begin{center}
		\centering{\includegraphics{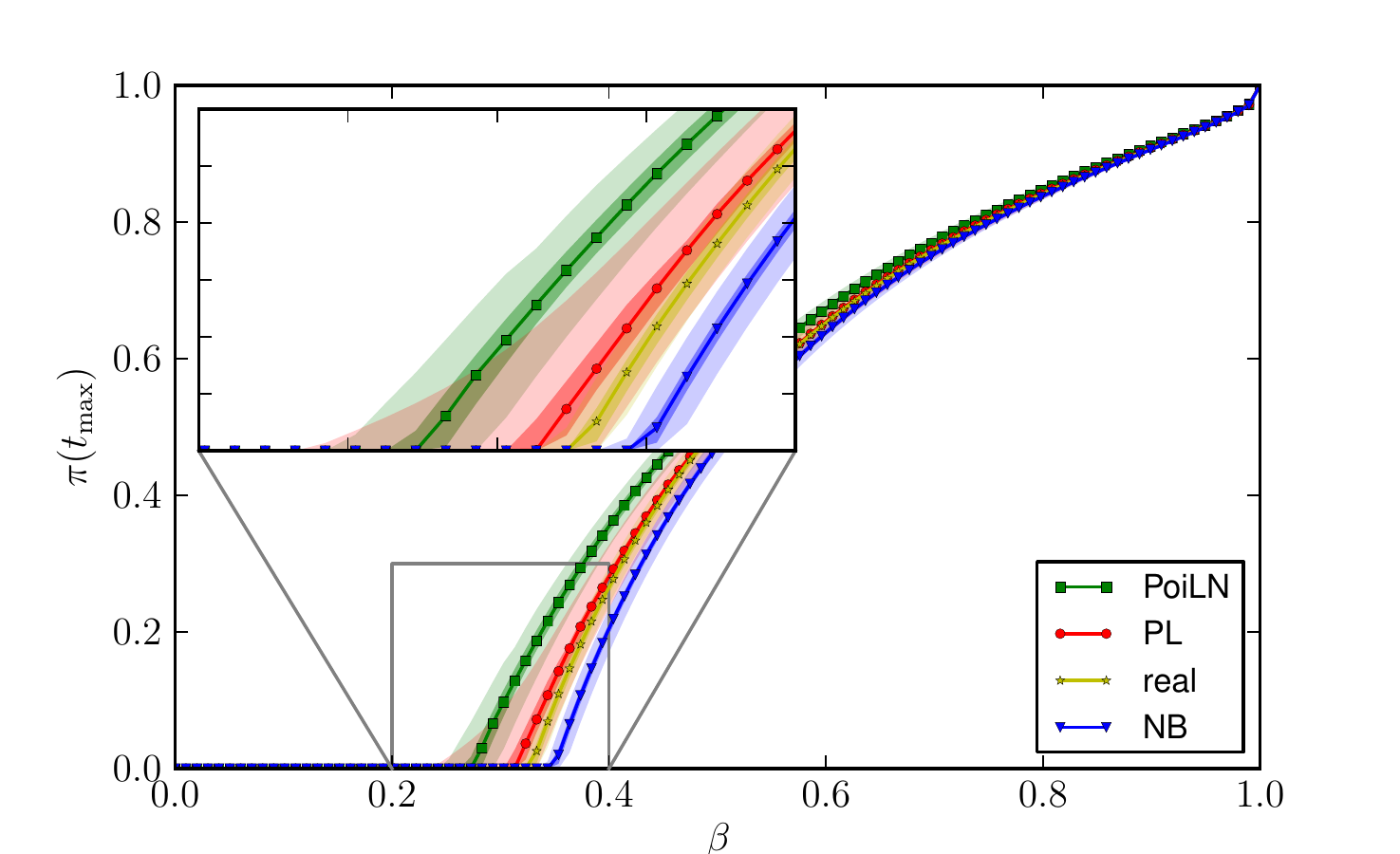}}
\end{center}
\caption{{\bf Median (lines), interquartile (darker areas) and $95\%$ confidence intervals (lighter areas) of the final prevalence as a function of the transmission probability.} $f$, fraction of nymphs among ticks is fixed to $10\%$.}
\label{prev_per10} 
\end{figure}

In this section we explored the effect of different fractions, $f$, of nymphs overall the total number of ticks on epidemic spreading. By exploring $f=2\%$ in main text, $f=5\%$ in Figure~\ref{prev_per5} and $f=10\%$ in Figure~\ref{prev_per10} we conclude that the larger the $f$, and the larger the probability of invasion of the pathogen. Moreover, it is worth to stress out that for different values of $f$ epidemic curves qualitatively do not change and in particular the order of epidemic thresholds is maintained.

\end{document}